\documentclass{tMOP2e}

\begin{document}

\title{Estimating the plasmonic field enhancement using high-order harmonic
generation: The role of inhomogeneity of the fields}
\author{T. Shaaran$^{a}$$^{\ast}$\thanks{$^\ast$Corresponding author. Email:
tahir.shaaran@icfo.es \vspace{6pt}} , M. F. Ciappina$^{a}$ and M. Lewenstein$%
^{a,b}$ \\
\vspace{6pt} $^{a}$\emph{ICFO-Institut de Ci\`ences Fot\`oniques,
Mediterranean Technology Park, 08860 Castelldefels (Barcelona), Spain};\\
$^{b}$\emph{{ICREA-Instituci\'o Catalana de Recerca i Estudis Avan\c{c}ats,
Lluis Companys 23, 08010 Barcelona, Spain}}\\
\vspace{6pt}}
\maketitle

\begin{abstract}
In strong field laser physics it is a common practice to use the high-order
harmonic cutoff to estimate the laser intensity of the pulse that generates
the harmonic radiation. Based on the semiclassical arguments it is possible to
find a direct relationship between the maximum value of the photon energy
and the laser intensity. This approach is only valid if the electric field driving HHG
is spatially homogenous. In laser-matter processes driven by plasmonics fields, the
enhanced fields present a spatial dependence that strongly modifies the electron motion and consequently the
laser driven phenomena. As a result, this method should be revised in order to more realistically estimate the field.
In this work, we demonstrate how the inhomogeneity of the fields will effect this estimation.
Furthermore, by employing both quantum mechanical and classical calculations, we show how one can obtain a
better estimation for the intensity of the enhanced field in plasmonic nanostructure.
 \bigskip

\end{abstract}






\section{Introduction}

Nowadays, there is a high demand for coherent light sources in the
ultraviolet (UV) to extreme ultraviolet (XUV) spectral range. These sources
are the essential ingredients for basic research, material science, biology
and recently lithography~\cite{misharmp}. Currently, high order harmonic generation
(HHG) is the well-known and experimentally proven method for producing coherent XUV. In order to generate high order harmonic, we need to amplify the output of the current femtosecond oscillators by carrying out complex processes like chirped-pulse amplification. In addition, the XUV based on HHG has low duty cycle and efficiency \cite{kim}.

 \ \

In the case using plasmonic enhanced fields to generate high-order
harmonics, it is not necessary to utilize extra cavities or laser pumping to
amplify the pulse power. Indeed, by exploiting surface plasmon resonances,
the local electric fields can be enhanced by more than 20dB~\cite%
{muhl,schuck}. In here, the input driven laser field is largly amplified to reach the threshold intensity required for high-order harmonic generation
(HHG) in noble gases. In addition, the pulse repetition rate remains
unchanged without employing any extra pumping or cavities. Furthermore, the
high harmonics radiation generated from each nanostructure acts as a
point-like source, which through constructive interference will provide even
more focused coherent radiations. This will provide a wide range of
possibilities to spatially rearrange the nanostructures to enhance and shape
the HHG spectral.

\ \

The basic principle of HHG based on plasmonics can be described as follows (~%
\cite{kim}). As a result of coupling of the external femtosecond low
intensity pulse with the plasmon mode, a collective oscillation of free
charges is induced within the localized regions of the nanostructure. The
free charges redistribute the electric field around the nanostructure
vicinity and form a spot of highly enhanced electric field. The enhanced
field exceeds the threshold intensity needed for generating high order
harmonic. Now, by injecting noble gases into the spot of the enhanced field
HHG can be produced.

\ \

In the experiment of Kim et al.~\cite{kim}, which has been recently under an intense scrutiny, (Ref. [15-17] of rapid), the
output of modest femotosecond oscillator (a pulse with 10 fs duration, 800
nm wavelength and 10$^{11}$ W/cm$^{2}$ intensity) was directly focused into
bow-tie nanoantenna array filled by argon gas to produce XUV radiations.
Despite using such low intensity, which is two orders of magnitude less than
the threshold intensity required for generating HHG in noble gases, they managed to
generate XUV wavelengths from the 7th (114 nm) to the 17th (47 nm)
harmonics. Based on finite-element calculation, they showed the laser pulse intensity was enhanced by a factor of 2-4 order of magnitude, which is large enough to produce harmonic far beyond 17th. It is clear that there is a very big difference between the theoretical and experimental estimations.

\ \

Due to the strong confinement of plasmonic hot spots the
locally enhanced field is spatially inhomogeneous, which will strongly influence
the subsequent motion of the electron in the continuum. As a result, the main features of strong field phenomena could be alerted in such non-homogenous field. Indeed, our previous works and on HHG \cite%
{ciappi2012,ciappiprl,tahirsfa} and above-threshold ionization(ATI)~\cite{ciappiati} demonstrated that the inhomogeneity strength of the field plays an important role for extending the cutoff of HHG and ATI. Ref.  \cite{yavuz,husakou} have reported similar behavior. Thus the local
field enhancement can only be realistically estimated if we take this effect
into the account. In fact, in Ref \cite{kim}, the harmonic order was used to
estimate the resonant plasmon field enhancement without taking this into the account. In this paper, we will investigate the relationship between the inhomogeneity strength of the field and HHG cutoff to estimate the field enhancement of the plasmonic nanostructure.

\ \

From theoretical point of view, the HHG process can be tackled using
different approaches (for details see the review articles in ~\cite%
{book1,book2}). In this article, we concentrate our effort in the estimation
of the laser intensity in nonhomogeneous fields using a well established
quantum mechanical approach, i.e. the Time Dependent Schr\"{o}dinger
Equation (TDSE) in reduced dimensions.

\ \

The paper is organized as follows. In the next section (Sec. II) we present
our theoretical approach to model high-order harmonic generation (HHG)
spectra produced by non-homogeneous fields. In Sec. III we employ the HHG
spectra calculated via the 1D-TDSE to estimate the laser intensity in the
experiment of Ref. \cite{kim}. We discuss how the nonhomogeneous character
of the laser electric field produces noticeable modifications in the HHG
spectra and how it is possible to reach the similar cutoffs by employing
smaller laser intensities. The paper ends with a short summary and an brief
outlook.

\section{Theory}

\indent We model the high-order harmonic generation (HHG) by considering the
dynamics of the free electron to be mainly along the
direction of the laser field, i.e. linearly polarized laser field. As a result,
it is reasonable to use the time dependent Schr\"odinger equation in
reduced dimensions((1D-TDSE)). It reads

\begin{eqnarray}  \label{tdse}
\mathrm{i} \frac{\partial \Psi(x,t)}{\partial t}&=&\mathcal{H}(t)\Psi(x,t) \\
&=&\left[-\frac{1}{2}\frac{\partial^{2}}{\partial x^{2}}%
+V_{atom}(x)+V_{laser}(x,t)\right]\Psi(x,t),  \nonumber
\end{eqnarray}

where $V_{atom}(x)$ and $V_{laser}(x,t)$ represent the atomic potential and the
potential due to the laser electric field, respectively. For $V_{atom}$ , we use
soft-core potential
\begin{eqnarray}  \label{atom}
V_{atom}(x)&=&-\frac{1}{\sqrt{x^2+a^2}}
\end{eqnarray}
which was first introduced in~\cite{eberly} and since then has been widely used
in laser-matter processes of atoms.
 In Eq. (\ref{atom}), the parameter $a$ is chosen in a such way to match the ionization potential of the atom under
consideration. In this work, we set $a=1.19$ in order to model argon atom.
The laser electric field is
linearly polarized along the $x$-axis and the potential associated is
modified to incorporate nonhomogeneous fields. Consequently we write
\begin{eqnarray}  \label{vlaser}
V_{laser}(x,t)&=&E(x,t)\,x
\end{eqnarray}
with laser electric field is given by
\begin{equation}  \label{electric}
E(x,t)=E_0\,f(t)\, (1+\varepsilon g(x))\,\sin\omega t.
\end{equation}
In (\ref{electric}), $E_0$ is the electric peak
amplitude and $\omega$ is the frequency of the coherent electromagnetic
radiation. $f(t)$ defines the pulse envelope and $\varepsilon$
is a small parameter that characterizes the inhomogeneity region. The $g(x)$
function represents the functional form of the nonhomogeneous field. For the plasmonic enhanced fields, to a first
approximation one can employ $g(x)=x$. For the problem we study in this paper, this is a good approximation as far as strong field is
concern since the excursion of the electron is very small and it sees a liner inhomogeneous field. To model short laser pulses, we shall
use a sin-squared envelope $f(t)$ given by
\begin{equation}
f(t)=\sin ^{2}\left( \frac{\omega t}{2n_{p}}\right),  \label{ft}
\end{equation}%
where $n_{p}$ and $\tau =2\pi n_{p}/\omega$ are the total number of cycles and the total time duration of
the pulse, respectively.


We used Crank-Nicolson
scheme~\cite{keitel} for solving Eq. (\ref{tdse}). In order to avoid spurious reflections from the
boundaries, at each time
step, we multiplied the total electronic wave function by a mask function of the form $cos^{1/8}$, which starts from the 2/3 of the grid and varies from 1 to 0~\cite{mask}.

Once the electronic state $\Psi(x,t)$ at the end of the laser
pulse from is found, we calculated the harmonic spectrum by
Fourier transforming the acceleration $a(t)$ of its active electron~\cite%
{schafer}, i.e.
\begin{equation}  \label{spec}
D(\omega)=\left| \frac{1}{\tau}\frac{1}{\omega^2}\int_{-\infty}^{\infty}%
\mathrm{d} t\mathrm{e}^{-\mathrm{i} \omega t}a(t)\right|^2.
\end{equation}
In Eq. (\ref{spec}), $a(t)$ was obtained by using the following commutator
relation
\begin{equation}  \label{accel1D}
a(t)=\frac{\mathrm{d}^{2}\langle x \rangle}{\mathrm{d} t^2}=-\langle \Psi(t)
| \left[ \mathcal{H}(t),\left[ \mathcal{H}(t),x\right]\right] | \Psi(t)
\rangle,
\end{equation}
where $\mathcal{H}(t)$ is the Hamiltonian defined in Eq. (\ref{tdse}).
The function $D(\omega)$ is so-called the dipole spectrum since it gives the spectral profile measured in HHG experiments.

\begin{figure}[tbp]
\begin{center}
\resizebox*{8cm}{!}{\includegraphics{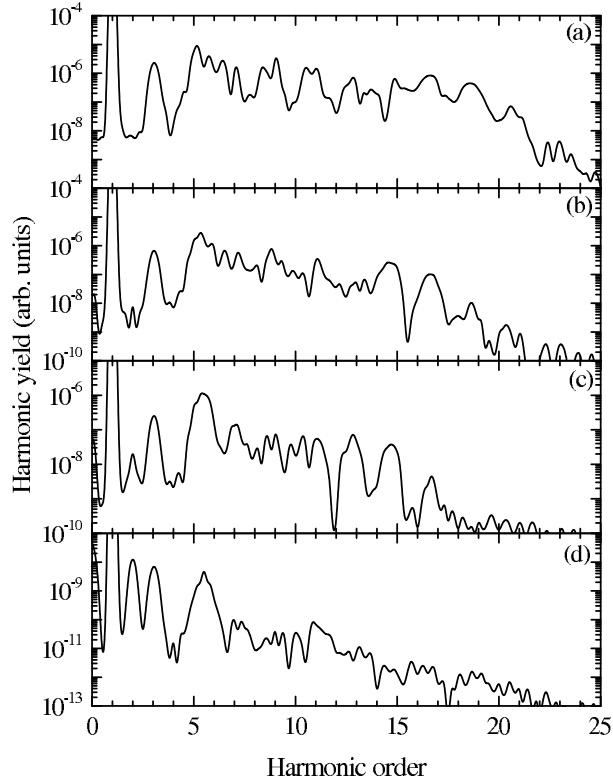}}%
%
\end{center}
\caption{HHG spectra calculated using the 1D-TDSE for Ar atoms. The laser
pulse has $\protect\lambda=800$ nm and a sin-squared envelope of 6 optical
cycles of total time duration. Panel (a) homogeneous case with laser
intensity $I=5.6\times10^{13}$ W/cm$^{2}$, panel (b) $\protect\varepsilon%
=0.001$ with laser intensity $I=4\times10^{13}$ W/cm$^{2}$, panel (c) $%
\protect\varepsilon=0.002$ with laser intensity $I=3\times10^{13}$ W/cm$^{2}$
and panel (d) $\protect\varepsilon=0.005$ with laser intensity $%
I=1\times10^{13}$ W/cm$^{2}$.}
\end{figure}

Using classical arguments it is possible to show that the harmonic radiation
has a cutoff that fulfill the relationship
\begin{equation}  \label{cutoff}
n_{c}\omega=3.17 U_p+I_p,
\end{equation}
where $n_c$ is the harmonic order at the cutoff \cite{sfa}. In here, $U_p$ denotes the ponderomotive energy defined as $U_p=I/4\omega^2$, with $I$ presenting the laser intensity in a.u.. From Eq. (\ref{cutoff}), it is possible to approximately
extract the field intensity using the experimental HHG spectra,
namely
\begin{equation}  \label{cutoffi}
I\,[\mathrm{a.u}]=\frac{4\omega^{2}}{3.17}(n_c\omega-|I_p|).
\end{equation}

In unit of W/cm$^{2}$, this leads  $I\,[\mathrm{W/cm}^{2}]=I_0 I\,[\mathrm{a.u}]$, where $I_0=3.51\times10^{16}$
W/cm$^{2}$ gives the atomic unit of intensity.

\ \

Eq. (\ref{cutoff}) is derived by solving the Newton equation of motion for an electron moving in a
oscillating field linearly polarized under the following
conditions (the three step or simple man's model~\cite{corkum}): (i) From position $x(t_i)=0$, the electron is ionized at time $t = t_i$ with zero velocity, then (ii) it starts accelerating in the field for half a cycle and when the electric field reverses its direction, (iii)returns to its initial position at a time $t=t_r$ (\textit{recollides} with its parent ion).
The electron kinetic energy at the return time $t_r$ can be obtained from $E_k(t_r)=\dot{x}(t_r)^{2}/2$. Eq.
By finding the value of $t_r$ (as a function of $t_r$) which maximizes this energy we will fulfill Eq. (\ref{cutoff}).

\ \

To estimate the intensity of the laser resonant enhanced field, one cannot simply use  Eqs. (\ref{cutoff}) and (\ref%
{cutoffi}). These relations are not anymore valid because the field is not spatially homogenous and one needs to take this
factor into the account in order to correctly estimate the field intensity. In our scheme, however, the
laser intensity can be extracted by inspecting the HHG spectra obtained from
the 1D-TDSE. The procedure would be as follows: (i) for a given cutoff $n_c$%
, make a first estimation of the laser intensity using Eq.(\ref{cutoffi}),
i.e. using an homogeneous electric field; (ii) introduce a small value of $%
\varepsilon$ and solve the 1D-TDSE by reducing the laser intensity (in general smaller
intensity values will be needed) till the same cutoff $%
n_c$ prescribed in (i) is reached; (iii) record the values of $\varepsilon$
and $I$ to establish if they are consistent with the geometry of the
nanostructure and the experimental conditions. %

\section{Results and discussion}

We employ the 1D-TDSE to calculate HHG spectra using the actual parameters
presented in Kim et al~\cite{kim}, a laser with wavelength of $\omega=0.057$ a.u. (800 nm)
and argon atom with potential of $I_p=0.579$ a.u. (15.7596 eV). From the experimentally measured spectra it is possible to observe a HHG cutoff at of  $n_c\approx 17$ (see Fig. 4 of Ref.~\cite{kim}). Using Eq. (\ref{cutoffi}), this cutoff corresponds to the plasmonic field intensity enhancement of $I\approx 5.6\times10^{13}$ W/cm$^{2}$. In order to validate this result, we use the obtained intensity and calculate HHG spectra by employing our 1D-TDSE model. For this simulation, we use a sin-squared shaped laser pulse with total time duration of 6 optical cycles (16 fs). Our calculations gives the same cutoff as the measurements of Kim et al.~\cite{kim}($n_c=17$), as shown in Fig. 1 (a).

\ \

\begin{figure}[tbp]
\begin{center}
\resizebox*{7cm}{!}{\includegraphics{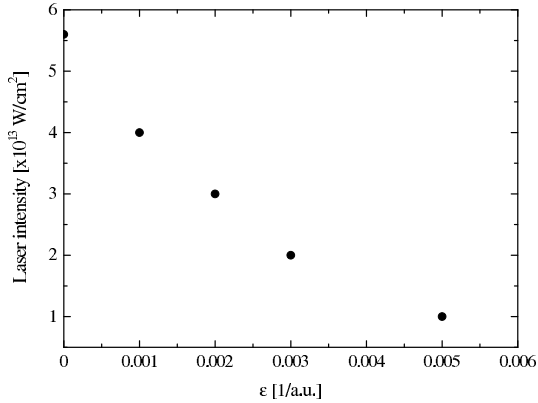}}%
%
\end{center}
\caption{Laser intensity as a function of the inhomogeneity parameter $%
\protect\varepsilon$ for a fixed value of the harmonic cutoff ($n_c=17$).}
\end{figure}

In the next step, we introduced small values of $\varepsilon$ in our model to demonstrate the behavior of the HHG spectra.
In this demonstration we keep the cutoff $n_c\approx 17$ constant and for a given $\varepsilon$ we reduce the intensity until  matching the cutoff.
In our model, which we considered a linear term to approximated the non-homogenous field, this parameter($\varepsilon$) characterizes the inhomogeneity region~\cite{husakou} which could be related to the gap size of the bow-tie nanostructure of Ref. \cite{kim}.

In panels (b), (c) and (d), we plotted HHG spectra for $%
\varepsilon=0.001$ and $I=4\times10^{13}$ W/cm$^{2}$, $\varepsilon=0.002$
and $I=3\times10^{13}$ W/cm$^{2}$ and $\varepsilon=0.005$ and $%
I=1\times10^{13}$ W/cm$^{2}$, respectively. The different panels correspond
to inhomogeneity regions ($\xi=1/\varepsilon$) of $\xi=1000$ a.u. (53 nm), $%
\xi=500$ a.u. (26.5 nm) and $\xi=200$ a.u. (10.6 nm), respectively. Based on our simulations, we could argue that if we allow
a small spatial inhomogeneity for the laser electric field ($<0.5 \%$), the
laser intensity in the center of the gap of the bow-tie nanostructure could have
values from two to five times smaller than the values obtained using Eq. (\ref{cutoff}).

 \ \

To summarize our calculations, in Fig. 2 we plot the laser intensity $I$
[W/cm$^2$] against the inhomogeneity parameter $\varepsilon$ for a fixed
value of the harmonic cutoff ($n_c=17$). From the figure it can be clearly seen that
by increasing the value $\varepsilon$ smaller values of laser
intensity $I$ are needed to reach the same HHG cutoff. This behavior is
related with the modifications in the electron trajectories introduced by
the non-homogeneous character of the field (for
details see~\cite{ciappi2012,ciappiprl,tahirsfa}).

\section{Conclusions}

In this work, we demonstrate how the inhomogeneity of the fields will effect the estimation of the field intensity from
 high-order harmonic generation spectra. We employing both quantum mechanical and classical calculations to demonstrate a method for
better estimation for the intensity of the enhanced field in plasmonic nanostructure. Our model, based on on the numerical solution of the time dependent Schr\"odinger equation (TDSE) in reduced dimensions, presents a large flexibility and robustness. In addition, due to the modest amount of
computational resources needed, it allows to perform numerous computations in a reasonable time.
Our findings show that by incorporating a very small inhomogeneity the laser intensity required to reach a determined
high order harmonic cutoff is substantially reduced.
The procedure is demonstrated in this work is suitable when, in the region where the electron dynamic takes place, the function form of the inhomogeneous field is linear. This is true of the electron excursion is small. However, for the case with large electron excursion(our work in !!)the situation is more complex and providing a better estimation for the intensity of the field is very challenging. Despite our developed 1D-TDSE model allows, in principle, to include any functional form for the nonhomogeneous fields to calculate HHG, extracting the intensity of the field by just considering HHG cutoff would not be easy.

\section*{Acknowledgments}

We acknowledge the financial support of the MINCIN projects (FIS2008-00784
TOQATA and Consolider Ingenio 2010 QOIT) (M. F. C. and M.L.); ERC Advanced
Grant QUAGATUA, Alexander von Humboldt Foundation and Hamburg Theory Prize
(M. L.); this research has been partially supported by Fundaci\'o Privada
Cellex. M.F.C. acknowledges Mitsuko Korobkin and Dane Austin for help and
advices in the numerical implementation of the 1D-TDSE model.

\end{document}